# A Relative Liutex Method for Vortex Identification

Jiawei Chen[a], Yifei Yu[b], Chaoqun Liu[b]*

[a]Department of Aerospace Engineering, Iowa State University, Ames, Iowa 50011, USA

[b]Department of Mathematics, University of Texas at Arlington, Arlington, Texas 76019, USA

* Correspondence: cliu@uta.edu

**Abstract:** A relative Liutex vortex identification method is proposed in this study, together with its explicit mathematical formulation. The method is designed to identify vortical structures based solely on local flow-field information and is inherently Galilean invariant, ensuring robustness under different reference frames. To validate the proposed approach, a three-dimensional flat-plate boundary-layer transition case is investigated, in which the relative Liutex is systematically compared with conventional vortex identification methods, including the Q-criterion and the original Liutex method. The results show that the relative Liutex is capable of simultaneously capturing both strong and weak vortical structures. Importantly, its behavior cannot be interpreted as a simple superposition of Liutex iso-surfaces obtained using different threshold values. Instead, the relative Liutex provides a more selective and physically coherent identification of weak vortices, particularly in regions above the Λ-vortex and in the downstream hairpin-vortex structures, while effectively suppressing spurious and noise-induced features. These advantages arise from its formulation based on local velocity-gradient strength rather than a global vortical-intensity measure. Owing to its ability to consistently identify vortices across a wide range of intensities, the relative Liutex demonstrates strong potential for revealing complex vortex structures and underlying flow mechanisms in vortex-dominated flows.

## 1. Introduction

Vortices are ubiquitous in nature across a wide range of scales, from microscopic motions to large-scale phenomena such as typhoons, hurricanes, cyclones and tornadoes, many of which exhibit strong destructive potential. More importantly, turbulent flows contain a large variety of vortical structures with different length scales and intensities, which play a crucial role in the generation, evolution, and maintenance of turbulence. Despite their fundamental importance, a universally accepted definition of a vortex has long been lacking[1][2].

The classical definition of a vortex is vorticity, which has a precise mathematical meaning as the curl of the velocity vector. As early as 1858, Helmholtz introduced the concept of a vortex filament[3] by considering a vorticity tube with an infinitesimal cross-section. Subsequently, Lamb adopted a simplified terminology in his classic monograph[4], referring to a vortex filament simply as a vortex. Because vorticity is a well-defined quantity, vorticity dynamics has been systematically developed to describe its generation and evolution, and has been widely applied to the investigation of vortical-flow stability and coherent vortical structures in transitional and turbulent flows[5][6][7]. However, the use of vorticity encounters fundamental limitations in viscous flows, particularly in turbulent regimes. First, vorticity alone cannot reliably distinguish genuine vortical structures from strong shear layers. Second, numerous studies have shown that in turbulent wall-bounded flows, the local vorticity vector is not necessarily aligned with the actual orientation of vortical structures, especially in near-wall regions[8][9][10]. Third, the location of maximum vorticity does not generally coincide with the core of a vortical structure[11][12].

The challenges in identifying and visualizing vortical structures in turbulent flows have motivated the rapid development of various vortex identification methods. These Eulerian vortex identification criteria are predominantly based on the analysis of the velocity gradient tensor, with most being exclusively determined by the eigenvalues of the velocity gradient tensor or related invariants, thus classified as eigenvalue-based criteria. The $\Delta$ criterion[13][14][15], proposed in 1983, emerged as one of the earliest eigenvalue-based methods utilizing tensor invariants for vortex identification. Subsequently, the Q criterion[16] was introduced in 1988, establishing itself as another widely adopted eigenvalue-based approach that measures the local balance between rotation rate and strain rate. The $\lambda_2$ criterion[17], developed in 1995, distinguishes itself by identifying vortices through the second-largest eigenvalue of the symmetric tensor derived from velocity gradient decomposition, though it cannot be directly expressed in terms of the eigenvalues of the velocity gradient tensor itself. The $\lambda_{ci}$ criterion[18][19], proposed in 1999, characterizes vortices using the imaginary part of the complex eigenvalues of the velocity gradient tensor; however, it has been found to suffer from serious contamination by shearing motion. More recently, the $\Omega$ method[20] was introduced in 2016 as a normalized measure of vorticity magnitude relative to total deformation. Despite their widespread use, these scalar-valued eigenvalue-based criteria share two fundamental limitations: the inadequacy in identifying the swirl axis or orientation, and the contamination by shearing motion for criteria associated with complex eigenvalues.

Beyond the vortex identification methods introduced above, a more advanced approach is the Liutex-based vortex identification method and its variants. Liu et al.[21] pointed out that vorticity alone cannot accurately represent the local rigid-body rotation of fluid elements, and that it should be further decomposed into rotational and non-rotational components. Based on this insight, the Liutex vector[22] was proposed to extract the rigid rotation embedded in the local fluid motion. Accordingly, the vorticity vector can be decomposed into a rigid rotation component and an antisymmetric shear component. Similarly, the velocity gradient tensor can be decomposed into a rigid rotation part and a non-rotation part. This decomposition is fundamentally different from the classical Cauchy–Stokes decomposition, which separates velocity gradient tensor into symmetric and antisymmetric parts but does not isolate the true rigid rotation of the local flow. Based on these developments, a relatively complete theoretical framework[22][23][24] for Liutex-based vortex identification has been established.

In this paper, a relative Liutex vortex identification method is proposed, together with its explicit mathematical formulation. The method is designed to identify vortical structures based solely on local velocity-gradient information and is inherently Galilean invariant. To assess its effectiveness, the proposed relative Liutex is applied to a three-dimensional (3D) flat-plate boundary-layer transition case and systematically compared with conventional vortex identification approaches, including the Q-criterion and the original Liutex method. Particular attention is paid to the identification of weak vortical structures associated with Λ-shaped and hairpin vortices, as well as the suppression of spurious noisy structures commonly observed in threshold-based methods.

The remainder of this paper is organized as follows. Section 2 introduces the theoretical formulation of the relative Liutex. Section 3 presents the numerical setup and comparative results. Concluding remarks and perspectives for future applications are given in Section 4.

## 2. Methodology

This section first introduces the Liutex vortex identification methodology, followed by the formulation of the relative Liutex concept based on triple decomposition of the velocity gradient tensor.

## 2.1 Liutex

The Liutex vector[22] $\boldsymbol{R}$ is an advanced vortex identification method developed to characterize the local rigid-body rotation of fluid elements. Unlike scalar vortex identification criteria such as $\Delta$ criterion[13][14][15], Q criterion[16], $\lambda_2$ criterion[17] and $\lambda_{ci}$ criterion[18][19], Liutex is a vector quantity, providing both the direction and strength of local rotation.

Specifically, the direction of $\boldsymbol{R}$ indicates the local axis of rotation, while its magnitude equals twice the local angular velocity associated with rigid-body rotation. Liutex was developed to overcome the limitations of vorticity-based measures, which cannot distinguish rigid rotation from shear contamination. Kolář and Šístek[25] have demonstrated that R is unaffected by stretching and shear, confirming its robustness for vortex identification and analysis.

The explicit expression[26] of the Liutex vector is given by

$$\boldsymbol{R} = R\boldsymbol{r} = \left[ \boldsymbol{\omega} \cdot \boldsymbol{r} - \sqrt{(\boldsymbol{\omega} \cdot \boldsymbol{r})^2 - 4\lambda_{ci}^2} \right] \boldsymbol{r} \qquad (1)$$

Where $\boldsymbol{\omega}$ is the vorticity vector, $\lambda_{ci}$ is the imaginary part of the complex conjugate eigenvalue of the velocity gradient tensor, and $\boldsymbol{r}$ is the eigenvector corresponding to the real eigenvalue.

## 2.2 Relative Liutex

In this study, a relative Liutex vortex identification method is introduced. This approach is derived from the triple decomposition[23][24][27] of the velocity gradient tensor (VGT), which provides a more refined description of local flow kinematics than the conventional symmetric–antisymmetric decomposition.

Traditionally, the velocity gradient tensor $G_{ij} = \nabla \vec{V}$ (VGT) is decomposed into a symmetric strain-rate tensor and an antisymmetric vorticity tensor. However, this decomposition cannot clearly separate shear deformation from rigid-body rotation. The triple decomposition framework addresses this limitation by partitioning the VGT into three physically distinct components: normal straining, pure shearing, and rigid-body rotation. The VGT can be expressed in its principal reference frame. In this frame, the VGT is quasi-triangular, and it can be decomposed as:

$$G_{ij} = G_{ij}^N + G_{ij}^R + G_{ij}^S = \begin{bmatrix} \varepsilon_1 & 0 & 0 \\ 0 & \varepsilon_2 & 0 \\ 0 & 0 & \varepsilon_3 \end{bmatrix} + \begin{bmatrix} 0 & 0 & 0 \\ 0 & 0 & \varphi_1 \\ 0 & -\varphi_1 & 0 \end{bmatrix} + \begin{bmatrix} 0 & \gamma_3 & \gamma_2 \\ 0 & 0 & \gamma_1 \\ 0 & 0 & 0 \end{bmatrix} \quad (2)$$

where $G_{ij}^N$, $G_{ij}^R$, and $G_{ij}^S$ represent the normal straining, rigid-body rotation, and pure shearing tensors, respectively. These tensors can be determined and transformed back to the original coordinate system using the ordered real Schur decomposition[28] of $G_{ij}$.

The overall strength of the velocity gradients can be expressed as

$$G_{ij}G_{ij} = G_{ij}^N G_{ij}^N + G_{ij}^R G_{ij}^R + G_{ij}^S G_{ij}^S + G_{ij}^R G_{ij}^S \quad (3)$$

where the first three terms represent the strengths of the individual components, and the last term accounts for the interaction between rigid-body rotation and shear.

Based on this formulation, the relative contribution of normal straining is $gg_{ns} = G_{ij}^N G_{ij}^N / G_{ij}G_{ij}$, the relative contribution of rigid rotation is $gg_{rr} = G_{ij}^R G_{ij}^R / G_{ij}G_{ij}$, the relative contribution of pure shearing is $gg_{ps} = G_{ij}^S G_{ij}^S / G_{ij}G_{ij}$, and the relative contribution of the interaction between shearing and rigid is $gg_{rs} = G_{ij}^R G_{ij}^S / G_{ij}G_{ij}$. Notably, Chen et al.[29] employed these relative contributions to elucidate the underlying mechanisms of complex separated flows over iced swept wings. Their analysis highlighted the role of relative rigid rotation, demonstrating the potential of this approach as an effective vortex identification method.

The relative Liutex ($R^*$) is defined as a quantitative measure of the relative contribution of rigid-body rotation to the total velocity gradient. It is expressed as

$$R^* = \frac{G_{ij}^R G_{ij}^R}{G_{ij}G_{ij}} = \frac{R^2}{2tr\left[(\nabla \vec{V})^T (\nabla \vec{V})\right]} \quad (4)$$

It should be emphasized that the relative Liutex method is Galilean invariant, similar to the original Liutex method. This property is crucial for vortex identification. If a vortex detection method relies on quantities that are not Galilean invariant, such as streamlines[30] or pathlines[31], the identified vortex structures may change or even disappear when observed from different inertial reference frames. Such behavior contradicts physical intuition, as vortices are objective flow structures and should exist independently of the observer's frame of reference.

## 3. Results and discussions

The relative Liutex $R^*$ is employed to identify vortex structures in a three-dimensional flat-plate boundary-layer transition. Its performance is assessed through systematic comparisons with conventional vortex identification methods in order to demonstrate its advantages. The three-dimensional transitional boundary-layer flow over a flat plate is obtained from direct numerical simulation (DNS), consisting of approximately $6.0 \times 10^7$ grid points and more than 400,000 time steps, at a freestream Mach number of 0.5.

Figure 1 presents a comparison of three vortex identification methods—Q-criterion, Liutex, and relative Liutex—based on their corresponding iso-surfaces. Notable differences are observed between the relative Liutex method and the other two approaches. Both the Q-criterion and Liutex methods rely strongly on user-defined threshold values (with $Q = 0.01$ and $R = 0.06$, respectively). As illustrated in Figures. 1(a) and 1(b), these methods successfully identify the $\Lambda$ vortex and the downstream hairpin vortices, but fail to reveal the weaker spanwise vortices. In contrast, the relative Liutex method is capable of simultaneously identifying vortical structures across a wide range of strengths. As shown in Figure 1(c), it clearly captures the spanwise vortices, the $\Lambda$ vortex, and the downstream hairpin vortices within a single threshold value $R^* = 0.06$, demonstrating its superior sensitivity to both strong and weak vortical structures.

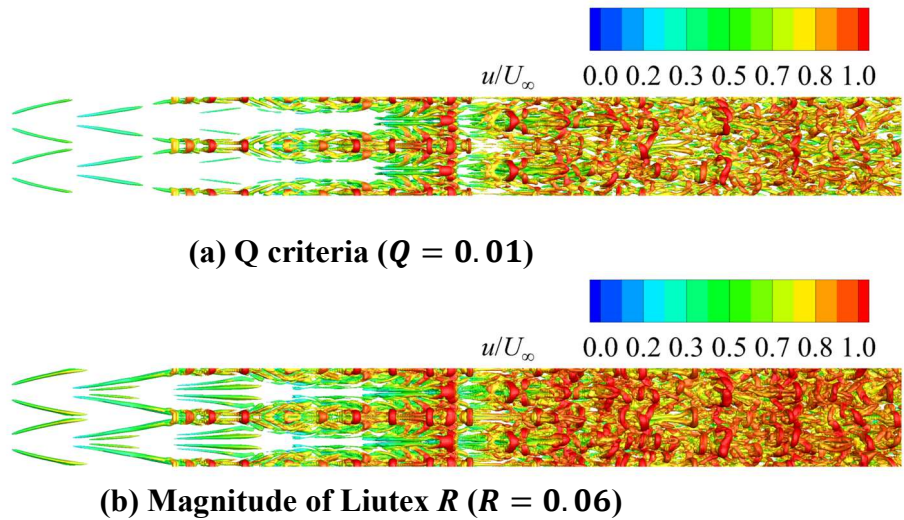

**(a) Q criteria ($Q = 0.01$)**

**(b) Magnitude of Liutex $R$ ($R = 0.06$)**

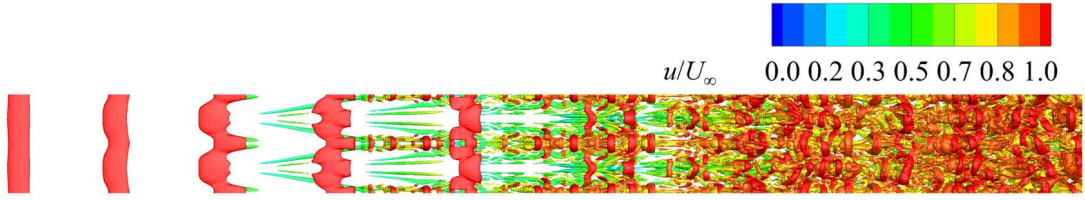

(c) Relative Liutex ($R^* = 0.06$)

Fig. 1. Comparison of three vortex identification methods—Q-criterion, Liutex, and relative Liutex—based on their corresponding iso-surfaces.

Figure 2 shows the visualization of weak spanwise vortices using Liutex iso-surfaces at low threshold values (R = 0.002–0.005). Compared with Figure 1(c), noticeable differences appear in the vortex structures when smaller Liutex thresholds are applied. This observation indicates that the relative Liutex method cannot be interpreted as a simple superposition of Liutex iso-surfaces at low and high threshold levels.

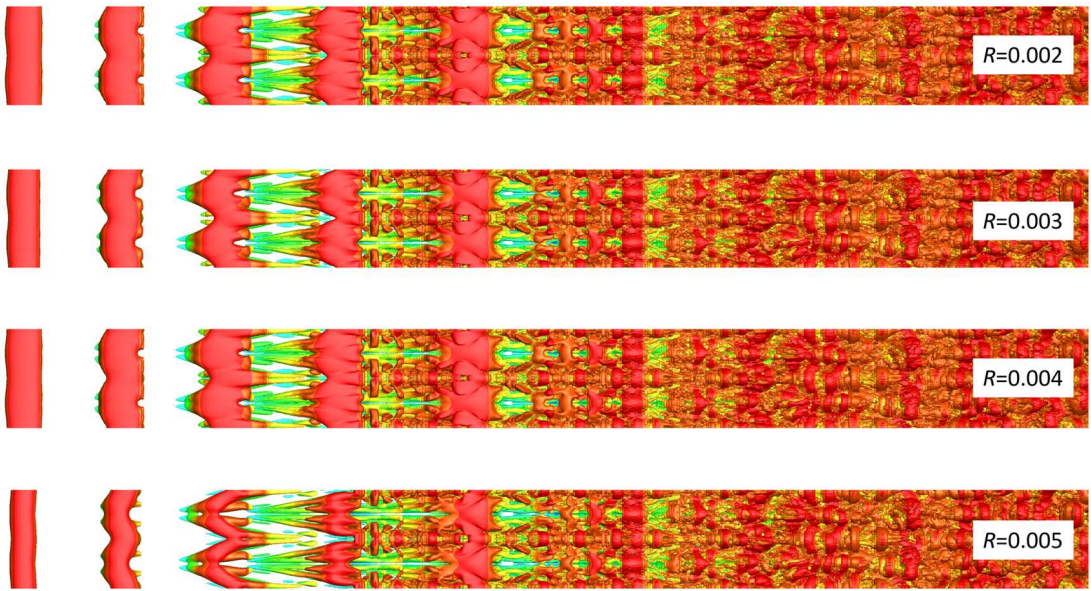

Fig. 2. Visualization of weak spanwise vortices using Liutex iso-surfaces at low threshold values ($R$ = 0.002–0.005).

Figure 3 further compares the 3D vortex structures identified by Liutex $R$ and relative Liutex $R^*$. In Figure 3(a), a relatively small threshold value of $R = 0.004$ is adopted to emphasize weak vortical structures. As a result, weak vortex structures are observed above the Λ vortex. Due to the low threshold, the downstream hairpin vortices are strongly contaminated by other weak vortical structures, leading to a cluttered flow topology. Figure 3(b) also reveals weak vortex structures above the Λ vortex; however, their spatial organization and morphology differ from those in Figure 3(a). Further downstream, the hairpin vortices exhibit a much cleaner structure, with significantly fewer weak vortices compared to Figure 3(a).

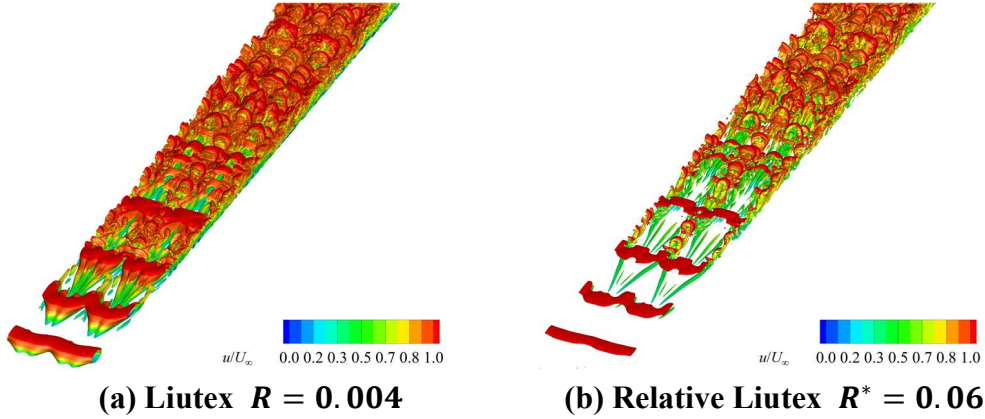

**(a) Liutex $R = 0.004$**       **(b) Relative Liutex $R^* = 0.06$**

**Fig. 3. Comparison of 3D vortex structures identified by Liutex and relative Liutex.**

The relative Liutex should not be interpreted as a simple superposition of Liutex iso-surfaces obtained using low and high threshold values. Its capability in identifying weak vortical structures differs fundamentally from that of the original Liutex method with a low threshold, and this difference is manifested in two distinct aspects. First, in the region above the Λ vortex, the relative Liutex captures weak vortical structures that are different from those identified by the Liutex method. Although both approaches are able to detect weak vortices in this region, the extracted structures exhibit discrepancies in their spatial organization and coherence. Second, when the threshold value of the Liutex method is further reduced, a larger number of weak vortical structures appear in the downstream hairpin-vortex region, leading to a cluttered visualization. In contrast, the relative Liutex is able to identify weak vortices in the same region while yielding significantly fewer structures, resulting in a much cleaner and more interpretable vortical field. This difference originates from the underlying definition of the two methods. The relative Liutex identifies weak vortices based on the local strength of the velocity gradient, whereas the original Liutex method effectively relies on a global measure of vortical intensity across the entire flow field. As a result, the relative Liutex provides a more selective and physically meaningful representation of weak vortical structures.

## 4. Conclusions

This study proposes a relative Liutex vortex identification method and presents its explicit mathematical formulation. To validate the proposed method, a three-dimensional flat-plate boundary-layer transition case is employed, in which the relative Liutex is systematically compared with conventional vortex identification approaches, including the Q-criterion and the original Liutex method. Several important conclusions can be drawn from this work.

(1) Relative Liutex is a new mathematical definition for the relative rigid rotation strength of the fluid motion, which is local, accurate, unique, systematical and Galilean invariant. Relative Liutex is an accurate definition for the rigid rotation part of fluid motion and a promising tool to identify vortices.

(2) The relative Liutex captures both strong and weak vortices simultaneously, but it is not equivalent to a simple superposition of Liutex iso-surfaces at different threshold levels. Relative Liutex captures weak vortices above the $\Lambda$-shaped vortex with improved spatial coherence and avoids excessive noisy structures in the downstream hairpin-vortex region. This advantage arises because the relative Liutex is defined based on local velocity-gradient strength rather than a global vortical-intensity measure, leading to a cleaner and more physically meaningful representation of weak vortical structures.

Future work will benefit from applying the relative Liutex method to a broader range of vortex-dominated flows. Owing to its capability to simultaneously identify both strong and weak vortex structures, the proposed method shows strong potential for revealing new vortex structures and underlying flow mechanisms.

## Acknowledgments

This work is supported by the National Science Foundation, with Grant No. 2422573. The authors also thankful to TACC for providing computation resources.

## Author declarations

**Conflicts of Interest**

The authors have no conflicts to disclose.

## Data availability

The data that support the findings of this study are available from the corresponding author upon reasonable request.